\documentclass[apreprint, 12pt]{aastex}
\usepackage{geometry}
\usepackage{natbib}

\newcommand{\xray}{\mbox{X-ray}}

\newcommand{\erg}{\mbox{erg}}

\newcommand{\persec}{\mbox{s$^{-1}$}}
\newcommand{\percmsq}{$cm^{-2}$}
\newcommand{\cgsflux}{\erg\,\percmsq\,\persec}

\def\simlt{\mathrel{\hbox{\rlap{\hbox{\lower4pt\hbox{$\sim$}}}\hbox{$<$}}}}
\def\simgt{\mathrel{\hbox{\rlap{\hbox{\lower4pt\hbox{$\sim$}}}\hbox{$>$}}}}
\def\ale{\mathrel{\hbox{\rlap{\hbox{\lower4pt\hbox{$\sim$}}}\hbox{$<$}}}}
\def\age{\mathrel{\hbox{\rlap{\hbox{\lower4pt\hbox{$\sim$}}}\hbox{$>$}}}}

\title{Early radio and X-ray observations of the youngest nearby type Ia supernova
  PTF\,11kly (SN\,2011fe) }

\author{
Assaf Horesh\altaffilmark{1}, 
S.~R.~Kulkarni\altaffilmark{1},
Derek B. Fox\altaffilmark{2}, 
John Carpenter\altaffilmark{1}, 
Mansi M. Kasliwal\altaffilmark{1,3}, 
Eran O. Ofek\altaffilmark{1,4},
Robert Quimby\altaffilmark{5},
Avishay Gal-Yam\altaffilmark{4},
S. Bradley Cenko\altaffilmark{6},
A.~G. de Bruyn\altaffilmark{7,8},
Atish Kamble\altaffilmark{9},   
Ralph A.~M.~J. Wijers\altaffilmark{9},
Alexander J. van der Horst\altaffilmark{10},  
Chryssa Kouveliotou\altaffilmark{11},
Philipp Podsiadlowski\altaffilmark{12},
Mark Sullivan\altaffilmark{12}, 
Kate Maguire\altaffilmark{12},
D. Andrew Howell\altaffilmark{13,14},
Peter E. Nugent\altaffilmark{15},
Neil Gehrels\altaffilmark{16},
Nicholas M. Law\altaffilmark{17},
Dovi Poznanski\altaffilmark{18} 
\&\
Michael Shara\altaffilmark{19}
}

\altaffiltext{1}{Cahill Center for Astrophysics, California Institute of Technology, Pasadena, CA, 91125, USA}
\altaffiltext{2}{Astronomy and Astrophysics, Eberly College of Science, The Pennsylvania State University, University Park, PA 16802, USA}
\altaffiltext{3}{Carnegie Institution for Science, 813 Santa Barbara St, Pasadena, CA, 91101, USA}
\altaffiltext{4}{Benoziyo Center for Astrophysics, Faculty of Physics,
  The Weizmann Institute of Science, Rehovot 76100, Israel}
\altaffiltext{5}{IPMU, University of Tokyo, Kashiwanoha 5-1-5, Kashiwa-shi, Chiba, Japan}
\altaffiltext{6}{Department of Astronomy, University of California, Berkeley, CA 94720-3411, USA}
\altaffiltext{7}{Netherlands Institute for Radio Astronomy (ASTRON), Postbus 2, 7990 AA Dwingeloo, The Netherlands} 
\altaffiltext{8}{Kapteyn Astronomical Institute, University of Groningen, Postbus 800, 9700 AA, Groningen, The Netherlands}
\altaffiltext{9}{
Center for Gravitation and Cosmology, University of Wisconsin, Milwaukee, 53211, WI}
\altaffiltext{10}{Universities Space Research Association, NSSTC, Huntsville, AL 35805, USA}
\altaffiltext{11}{Space Science Office, VP-62, NASA/Marshall Space Flight Center, Huntsville, AL 35805, USA}
\altaffiltext{12}{Department of Physics (Astrophysics), University of Oxford, Keble Road, Oxford OX1 3RH, UK}
\altaffiltext{13}{Las Cumbres Observatory Global Telescope Network,
Santa Barbara, California 93117, USA}
\altaffiltext{14}{Department of Physics, University of California Santa Barbara, Santa Barbara, CA 93106, USA}
\altaffiltext{15}{Computational Cosmology Center, Lawrence Berkeley National Laboratory, 1 Cyclotron Road, Berkeley, CA 94720, USA}
\altaffiltext{16}{NASA-Goddard Space Flight Center, Greenbelt, MD 20771, USA}
\altaffiltext{17}{Dunlap Institute for Astronomy and Astrophysics,
  University of Toronto, 50 St. George Street, Toronto M5S 3H4,
  Ontario, Canada}
\altaffiltext{18}{School of Physics and Astronomy, Tel-Aviv University, Tel-Aviv 69978, Israel}
\altaffiltext{19}{Department of Astrophysics, American Museum of Natural History Central Park West \& 79th street, New York, NY 10024}

\begin{document} 

\begin{abstract}

On August 24 (UT) the Palomar Transient Factory (PTF) discovered PTF11kly
(SN 2011fe), the youngest and most nearby type Ia supernova (SN Ia) in 
decades. We followed this event up in the radio (centimeter and millimeter
bands) and X-ray
bands, starting about a day after the estimated explosion time. We present
our analysis of the radio and X-ray observations, yielding the tightest
constraints yet placed on the pre-explosion mass-loss rate from the 
progenitor system of this supernova. We find a robust limit of $\dot
M\simlt 10^{-8}(w/100\,{\rm km\,s}^{-1})\,M_{\odot}\,{\rm yr}^{-1}$ from sensitive X-ray
non-detections, as well as a similar limit from radio data, which
depends, however, on assumptions about
microphysical parameters. We discuss our results in the context of 
single-degenerate models for SNe Ia and find that our observations 
modestly disfavor symbiotic progenitor models involving a red giant
donor, but cannot constrain systems accreting from main-sequence
or sub-giant stars, including the popular supersoft channel. In view
of the proximity of PTF11kly and the sensitivity of our prompt observations
we would have to wait for a long time (decade or longer) in order to
more meaningfully probe the circumstellar matter of Ia supernovae.

\end{abstract}

\bibliographystyle{apj}

\section{Introduction}
\label{sec:Introduction}

Type Ia supernovae (SNe) have served as an exquisite probe of
cosmography (Riess et al. 1998, Perlmutter et al. 1999) \nocite{riess+98, perlmutter+99}.  As a result of this role, this class has been studied
in unprecedented depth and breadth.  Nonetheless the progenitors
of Ia supernovae remain enigmatic. According to common wisdom a
type Ia SN is due to the thermonuclear explosion of a white dwarf
with a mass approaching the Chandrasekhar limit~\citep{hoyle+fowler60}.
The progenitor question is then one of understanding how a white dwarf can
approach the Chandrasekhar mass (see review
by~\citealt{hillebrandt+niemeyer00}). In the single degenerate (SD)
model~\citep{whelan+iben73}, a WD accretes mass from its hydrogen-rich
star companion, reaches a mass close to the Chandrasekhar mass,
which is sufficient to ignite carbon, and explodes. In the double
degenerate (DD) model, a supernova results from the merger of two
WDs~(\citealt{iben+tutkov84,webbink84}; see also
\citealt{yungelson+livio00})
.

It has long been suggested that radio and X-ray observations of
type Ia SNe have the ability to provide diagnostics to distinguish
between these two models (Boffi \&\ Branch 1995, Eck et al. 1995, 
Panagia et al. 2006)\nocite{Boffi+branch95,boffi+95,eck+95,panagia+06}. 
In most variations of the SD model, the winds from the
donor star will enrich the circumstellar medium. The interaction of
the blast wave from the supernova with the circumstellar medium can
result in radio emission. In contrast, there is no expectation of
circumstellar medium and hence of radio emission in the DD model.

On UTC 2011 August 24.16 the Palomar Transient Factory (PTF;
\citealt{law+09, rau+09}) discovered PTF\,11kly, a rapidly rising
transient, in the nearby (distance, $d\approx 6.4$\,Mpc;
\citealt{shappee+stanek11}) galaxy Messier~101 (Nugent et al. 2011\nocite{Nugent11}).  Spectroscopy
undertaken at the Liverpool Telescope led to a plausible Ia
classification\footnote{at which point the event was rechristened
to SN\,2011fe by the Central Bureau for Astronomical Telegrams} 
and was soon confirmed by observations at the Lick 3-m telescope and
the TNG. 

The apparent extra-ordinary youth and the proximity\footnote{The closest type Ia SN previous to PTF\,11kly is
1986G at a distance of 5.5\,Mpc} of PTF\,11kly presents a
unique opportunity to sensitivity probe the circumstellar medium of a
type Ia supernova.  Therefore,  we immediately initiated (Gal-Yam et
al. 2011\nocite{Gal-Yam11}) observations with 
the {\it Swift} Observatory, 
the Combined Array for Research in
Millimeter-wave Astronomy (CARMA)\footnote{
Support for CARMA construction was derived from the states of California, Illinois, and Maryland, the James S. McDonnell Foundation, the Gordon and Betty Moore Foundation, the Kenneth T. and Eileen L. Norris Foundation, the University of Chicago, the Associates of the California Institute of Technology, and the National Science Foundation. Ongoing CARMA development and operations are supported by the National Science Foundation under a cooperative agreement, and by the CARMA partner universities.} and 
the  Expanded
Very Large Array (EVLA)\footnote{The EVLA is
operated by the National Radio Astronomy Observatory, a facility
of the National Science Foundation operated under cooperative
agreement by Associated Universities, Inc.}.
A few days later low frequency observations were undertaken 
at  Westerbork Synthesis Radio
Telescope (WSRT)\footnote{The Westerbork Synthesis Radio Telescope is operated by  ASTRON
(Netherlands Foundation for Radio Astronomy) with support from the
Netherlands Organization for Scientific Research (NWO).}. X-ray
observations were obtained with the {\it Swift} and Chandra observatories.

\section{The Observations}
\label{sec:Observations}

The early optical light curve of PTF11kly shows an extra-ordinary good fit to that expected
from an exploding star (flux proportional to exponential of square
of time). As a result the birth of the supernova can be accurately
timed to a fraction an hour: UT 2011 August 23.69 (Nugent et al. 2011\nocite{Nugent11}).
Our first observations at both radio and X-ray bands
were taken just over a day after
the explosion.

\subsection{Radio}

\begin{figure}
\centering
\includegraphics[scale=0.6,angle=0]{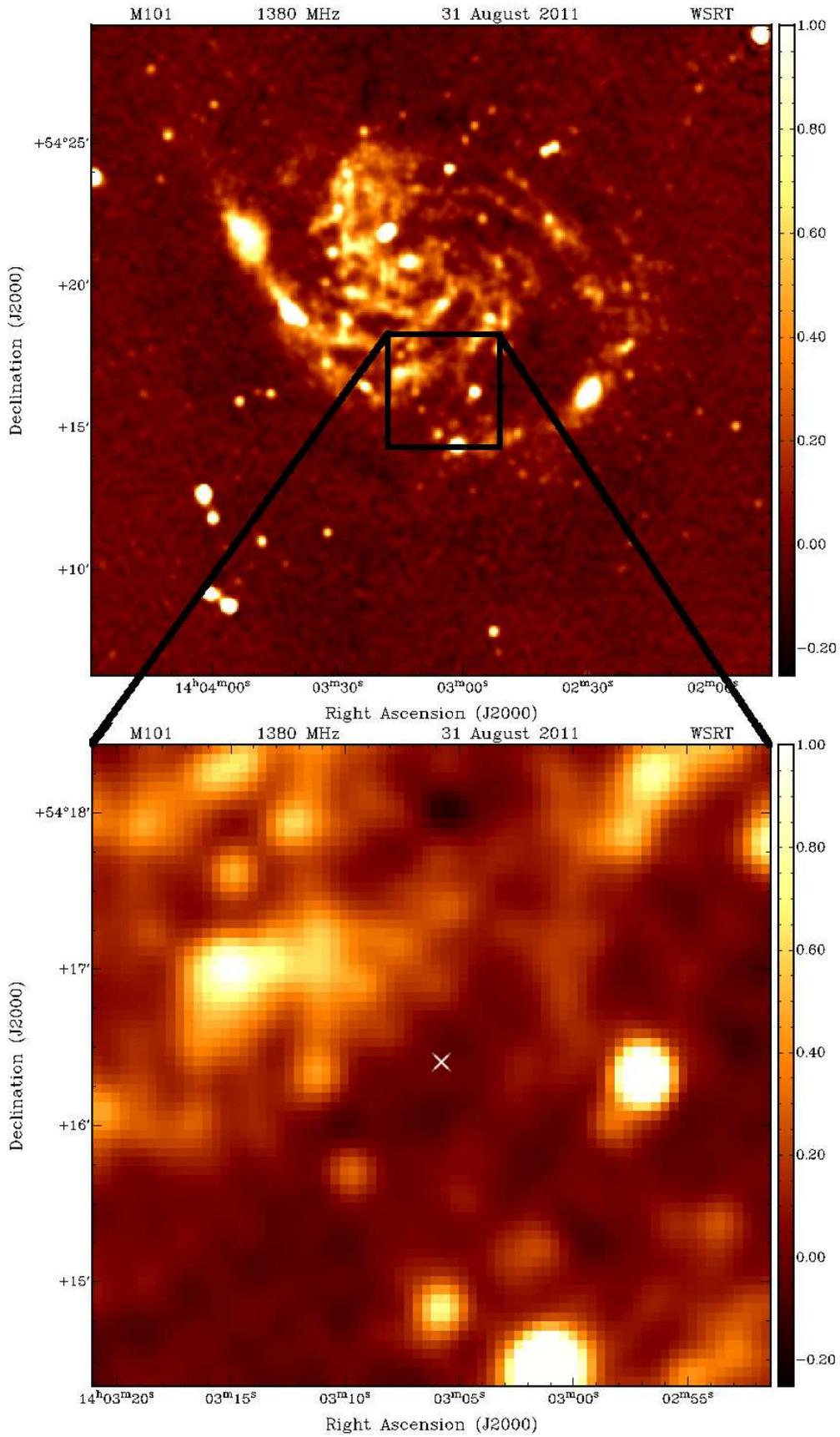}
\caption{21\,cm image of M101 taken with WSRT on UT 2011 August
  31. SN position is shown by cross.}
\label{fig:WSRTImage}
\end{figure}

The log of observations and the associated details can be found in
Table~\ref{tab:RadioLog}.  Our
CARMA and EVLA observations include the earliest search for radio
emission in cm-wave and mm-wave bands and subsequent observations
include a very sensitive search in the 21-cm band obtained at the
WSRT (see Figure~\ref{fig:WSRTImage}).  Following our first EVLA and
CARMA observations, additional data was taken at a lower frequency
(5 GHz) on UT 2011 August 25.8 by \citet{chomiuk+soderberg11}.  They
reported a null detection with  $-5\pm 6\,\mu$Jy. As can be gathered
from Table~\ref{tab:RadioLog} there are no detections at any epoch
and in any band. In the next section we discuss the implications
of these null detections.

\begin{table}
\caption{Log of radio observations}
\scalebox{0.9}{
\begin{tabular}{lcclccccc}
\hline\hline
Start & $\Delta T$ & $\tau$ & Facility & $\nu$ & $S_\nu$ & Luminosity 
 & $\dot M$ & Note\cr
 UT & day &  min   &             & GHz  & $\mu$Jy&
 $\simlt 10^{24}$\,erg\,s$^{-1}$\,Hz$^{-1}$ & $10^{-8}w_7/\epsilon_{-1}\,M_\odot\,{\rm yr}^{-1}$ 
 \cr
\hline
Aug 24.98  &  1.4 & 178 & CARMA & 93     & $-16\pm 510$ & 75 & 23 & (1) \cr
Aug 25.02  & 1.3  & 37 & EVLA     & 8.5  & $ -4.4 \pm 25$ & 3.7 & 1.7& (2) \cr
Aug 27.71  & 4.0  &  35 & EVLA      & 5.9     & $1.3\pm 7$ & 1.0 &1.1& (3) \cr
Aug 28.25  & 4.8  & 800  &WSRT     & 4.9   & $-71\pm 34$ & 5.0& 2.6 & (4) \cr
Aug 29.97  & 6.3  & 29  &EVLA      & 5.9     & $-0.9\pm 9$ & 1.3 & 3.1 & (3) \cr
Aug 31.38  &  8 & 630  & WSRT     &1.4 & $2\pm 25$ & 3.7 & 1.7 &
(4)\cr
\hline
\end{tabular}
}
\\
\\Notes:
{\small
The columns starting from left to right are as follows: start of
integration in UT; mean epoch of observation (in days since explosion);
integration time in minutes;
facility; central frequency in GHz; nominal flux and associated rms
in the vicinity of 
PTF11kly in $\mu$Jy; the corresponding 3-$\sigma$ spectral luminosity
assuming a distance of 6.4 Mpc to M101; inferred upper limit to the
mass loss rate (see \S\ref{sec:ConstraintsfromRadio} 
for explanation of parameters);  specific notes (see below).
Following Nugent et al. (2011) we assume that the explosion time
of PTF11kly is UT 2011 August 23.69. 
The following packages were employed to reduce the data
\texttt{AIPS} (EVLA), \texttt{Miriad} (CARMA) and 
\texttt{NEWSTAR} (WSRT).
(1) The CARMA observations
were obtained in the E configuration using only nine 6-m
antennas. The larger antennas were not available owing to reconfiguration
of the array. Bandwidth of 8000 MHz.
Calibrators: J1153+495 \&\ J1642+689 (phase) and MWC349 (flux). 
(2) Bandwidth of 256 MHz. Calibrators: J1419+5423 (phase) and
3C286 (flux). 
(3) Band width of 2000 MHz. Data obtained under Director's discretionary time
(PI: A. Soderberg). 
(4) Band width of $8\times 20$\,MHz. For each of the eight IF channels,
only 3/4 of the channel bandwidth  was used in
making the map. 
The flux density calibrators used were 3C147 and 3C286 on the
Baars et al. (1977)\nocite{Baars77} scale.
}
\label{tab:RadioLog}
\end{table}

\medskip
\noindent

\subsection{X-ray observation}
Following our classification of PTF11kly as a supernova we immediately 
triggered the {\it Swift} Observatory (see log of observations in
Table~\ref{tab:X-rayLog}) .
The first {\it Swift} observations\footnote{Target ID 32081, with
initial ObsID 32081001 lasting 4.5 ksec over three consecutive
orbits, followed by 53.7 ksec of observations through Sep 8.58
(ObsIDs 32081002--32081029).}  began on UT 2011 Aug 24.92. We used
the {\it Swift}-XRT data products generator of the UK {\it Swift}
Science Data Centre (SSDC; see \citealt{evans+09}) to generate a
single combined and astrometrically-corrected event file from the
first 58.2 ksec of exposure.  Six events are found within an aperture
of radius 9 XRT pixels (21.2\arcsec) centered on the position of
PTF11kly. This can be compared to independent background expectations
of $7.0\pm 0.4$ and $11.7\pm 0.7$ counts, respectively.  The
corresponding 90\%-confidence limits on total source counts and
aperture-corrected average count rate are $n_\gamma < 4.3$ and $r_X
< 0.1\times 10^{-3}\,{\rm s^{-1}}$, respectively.

However, we note that two of the counts within the source region
arrive within the first 2\,ksec -- an {\it a priori} unlikely
occurrence (at $\approx$90\% level of confidence) -- and could be
taken as evidence of early, bright \xray\ emission from PTF\,11kly,
at the $L_X\sim 10^{38}\,\rm{erg\,s}^{-1}$ level (see also Fox
2011a,b).  We realize that this statistical evidence does not warrant
a claim of detection. Nonetheless, given that the observations were
done at an extraordinarily early epoch of 1.21 day post-explosion
we think it worth highlighting this issue for the benefit of future
observers.

Our upper limit of $n_\gamma < 4.8$\,counts, from the first {\it Swift} observation sequence (4.5\,ksec exposure)
alone, provides the
following upper limits on X-ray flux (over the energy range
0.3--10\,keV)\footnote{ We assume a Galactic ISM contribution of
$N_H=1.8\times 10^{20}\,{\rm cm}^{-2}$; \citealt{kalberla+05}.
There may be an additional similar contribution from M101, depending
on the depth of the line of sight to the SN; Kamphuis et al.
(1991).\nocite{Kamphuis1991}}: $F_X< 6.2\times 10^{-14}$\,\cgsflux\
(thermal bremsstrahlung model with $kT=10$\,keV) and $F_X < 5.0\times
10^{-14}$\,\cgsflux\ (power law model with photon index, $\Gamma=2$).
Corresponding upper limits to the X-ray luminosity are $L_X< 3.0\times
10^{38}\,{\rm erg\,s}^{-1}$ and $L_X < 2.5\times 10^{38}\,{\rm
erg\,s}^{-1}$, respectively.

Observations\footnote{ObsID 14341; PI J. Hughes} with the Chandra
X-ray Observatory began on UT 2011 Aug 27.44 and lasted for 49.7\,ksec;
the mean epoch is UT 2011 Aug 27.74 (corresponding to 4\,days
post-explosion)(\citealt{Hughes+11}). No soft proton flaring was evident during the
observation. Only one event was found in the source region, a
2.5-arcsec aperture centered on the SN, whereas $2.23\pm 0.1$
background counts were expected (in an equivalent aperture).  The
90\%-confidence upper limit on the expectancy value of a Poisson
process that generates one count is 3.9 counts.  Ignoring the
background contribution, after applying the aperture correction (as
prescribed by Feigelson et al. 2002\nocite{feigelson+02}), we find
an upper limit to the count rate, $r_X\simlt 0.11\times 10^{-3}\,{\rm
 s}^{-1}$ (90\%-confidence).

For the thermal bremsstrahlung model this upper limit translates
to $F_X < 9.0\times 10^{-16}$\,\cgsflux\ (0.3--8.0 keV), corresponding
to $L_X < 4.4\times 10^{36}\,{\rm erg\,s}^{-1}$.  For the power law
model we find $F_X < 8.2\times 10^{-16}$\,\cgsflux\ (0.3--8.0 keV),
corresponding to $L_X < 4.0\times 10^{36}\,{\rm erg\,s}^{-1}$.

\begin{table}
\caption{Log of X-ray observations}
\scalebox{0.9}{
\begin{tabular}{lcclcccc}
\hline\hline
Start & $\Delta T$ & $\tau$ & Facility & Band & $F_X$ & Luminosity 
 & $\dot M$ \cr
 UT & day &  ksec  &             & keV  & $\simlt 10^{-16}$\cgsflux &
 $\simlt 10^{36}$\,erg\,s$^{-1}$ & $10^{-8}w_7/\epsilon_{e-1}\,M_\odot\,{\rm yr}^{-1}$ 
 \cr
\hline
Aug 24.92 & 1.21 & 4.5 & Swift & [0.3-10] & 500  & 250 & 20 \cr
Aug 27.44 & 4 & 49.7 & Chandra & [0.3-8] & 8.2  & 4 & 1.1 \cr
\hline
\end{tabular}
}
\\
\\Notes:
{\small
The columns starting from left to right are as follows: start of
integration in UT; mean epoch of observation (in days since explosion);
integration time in minutes;
facility; Energy band in keV; flux in $10^{-16}$\cgsflux; the corresponding luminosity
limit assuming a power law model with photon index $\Gamma=2$ and a distance of 6.4 Mpc to M101; inferred upper limit to the
mass loss rate (see \S\ref{sec:ConstraintsfromX-ray} 
for explanation of parameters).
}
\label{tab:X-rayLog}
\end{table}

\section{Mass-loss rate constraints}
\label{sec:MassLossRateConstraints}

The theory of  radio emission from SNe relevant to Type I events
is discussed in \citet{chevalier82, chevalier98} and
\citet{chevalier+fransson06} and a summary of radio observations
of supernovae can be found in \citet{weiler+02}. \citet{panagia+06}
report a comprehensive summary of searches for radio emission from Ia
supernovae.  

The basic physics is as follows.  The supernova shock-wave
ploughs through the circumstellar medium (CSM).  In the post-shock
layer, electrons are accelerated to relativistic speeds and strong
magnetic fields also appear to be generated. The relativistic
electrons then radiate in the radio via synchrotron emission.
X-rays are emitted via two energy channels: thermal bremsstrahlung
emission from hot post-shocked gas and inverse Compton 
scattering of the optical photons from the supernova by the relativistic
electrons. 

The density of the circumstellar medium is a key physical parameter.
After all a strong shock can only be generated if the SN blast wave can
run into CSM. Thus, the strength of the radio and X-ray emission
allows us to diagnose the CSM density.

For a star which has been losing matter at a constant rate,
$\dot M$, the circumstellar density has the following radial profile:
$\rho(r) = n(r)\mu= \dot M/(4\pi r^2 w)$ where $w$ is the wind
velocity, $r$ is the radial distance from the star, $n$ is the
particle density and $\mu$ is the mean atomic weight of the
circumstellar matter.  We assume that
the circumstellar matter is ionized (by the shock breakout and by
radiation from the young supernova).

\subsection{Constraints from Radio Measurements}
\label{sec:ConstraintsfromRadio}

Massive stars exploding as type Ib/Ic SNe with their fast moving
ejecta and low density CSM provide a good starting point to discuss
radio emission from Ia supernovae.  The spectrum of the radio
emission from type Ib/Ic events follows the ``Synchrotron
Self-Absorption'' (SSA) form: a $\nu^{5/2}$ low-frequency tail
(optically thick regime) and a declining power law, $\nu^\alpha$
at high frequency (optically thin regime). For most well studied
SNe $\alpha\approx -1$.  Diagnosis of the CSM is based on the peak
frequency ($\nu_m$; the synchrotron optical depth is unity at this
frequency) and the peak flux ($S_m$) and the evolution thereof.

We use the basic SSA formulation as in \citet{chevalier98} with some
modifications (noted below). We assume a fraction of electrons
are accelerated to relativistic energies and with a power law
distribution, $dN/dE=N_0E^{-p}$; here $E=\gamma m_ec^2$ is the
energy of electrons and $\gamma$ is the Lorentz factor. Electrons which
are not relativistic will not radiate strongly and thus we introduce
a minimum Lorentz factor, $\gamma_{\rm min}$. Only electrons with
$\gamma>\gamma_{\rm min}$ are assumed to contribute to the radio
emission.

As in all strong shocks, in addition to acceleration of electrons,
strong magnetic fields appear to be generated in the post shock
gas. Our current understanding is such that we simply parametrize
the relative energy fractions of each of these two components.

The thermal energy density of the post-shocked gas is $\frac{9}{8}\rho
v_s^2$ (where $v_s$ is the shock speed).  The magnetic energy density
is $B^2/(8\pi)$. We denote the ratio of this energy density to that of
the thermal energy density of post-shocked gas by $\epsilon_B$. We find
\begin{equation} u_{B}=\frac{B^{2}}{8\pi}\approx 6.8\times
	10^{9}m^{2}\epsilon_{B} \left(\frac{\dot M [M_{\odot} {\rm
	yr}^{-1}]}{w [{\rm km}~{\rm s}^{-1}]}\right)t_{\rm d}^{-2}
  {\rm erg\,cm}^{-3};
\label{eq:ub}
\end{equation}
here, $m$ is the power law exponent in the equation relating the radius
of the shock to time, $R_s\propto t^m$ and $t_d$ is the time post-explosion
in days.
In a similar manner, we let $\epsilon_e$ denote the ratio of the energy
density of relativistic electrons ($\gamma>\gamma_{\rm min}$) to the thermal
energy density. 

The use of radio diagnostics (unfortunately) involves the values of
several additional parameters. The first parameter is $p$, the power law index
of the relativistic electrons. Theory and observations of this value
agree that this value should be $p\approx 3$ (see \citealt{weiler+02, chevalier+fransson06}). Next, we
make the simplifying assumption that the blast wave moves at a
constant velocity, $v_s$; this is equivalent to setting $m=1$. 
This is admittedly a simplification\footnote{We are interested in
the very outer layers of the exploded star. Note that with (say) a mass
loss rate of $10^{-8}\,M_\odot\,{\rm yr}^{-1}$ and a wind velocity of $10^7\,{\rm cm\,s}^{-1}$
the leading edge of the shock ($v_{s}=4\times 10^{9} {\rm cm/s}$, say)
would have swept up a mere $10^{-8}\,M_\odot$ by
day 1. It is the velocity-density structure of such thin outermost
layer that determines $v_{s}$.}.
However, 
given our non-detections we felt that an analysis more sophisticated than this was
not warranted.  

Finally, we come to the microphysics parameters, $\epsilon_e$ and
$\epsilon_B$. It appears from well studied SN shocks that  
$\epsilon_e$ is usually about 0.1 with modest dispersion \citep{chevalier+fransson06}.
In contrast, $\epsilon_B$ appears to be a highly variable parameter.
This is quite understandable from simple considerations. The magnetic field
is generated both in the post-shock gas as well as by compression of 
the field already present in the pre-shock gas.  For instance a radio and
X-ray modeling of SN\,2002ap (a type Ic supernova with a presumed
Wolf-Rayet progenitor) by Bj\"ornsson \&\ Fransson (2004)\nocite{Bjornsson2004} find 
that the assumption of equipartition ($\epsilon_B=\epsilon_e$) lead to
inferring $\dot M$ much lower than the expected for a Wolf-Rayet progenitor.
The inferred mass loss rate can be increased if
$\epsilon_{\rm B}=
2\times 10^{-3}$. We find a similar imbalance between $\epsilon_B$ and
$\epsilon_e$  for SN\,2011dh (a type IIb
supernova; Horesh et al. 2011\nocite{Horesh2011}; see also Soderberg
et al. 2011\nocite{Soderberg+11}).
For this reason, unless stated otherwise, we will fix $\epsilon_e=0.1$
and assume that $\epsilon_B$ is a free parameter. Finally, since only
relativistic electrons contribute to radio and X-ray emission we adopt
$\gamma_{\rm min}=2$.

Following \citet{chevalier98}, the radio spectral luminosity in the
optically-thin regime is given by
\begin{equation}
	L_{\nu}=\frac{(4\pi)^{2}fR_{s}^{3}}{3}c_{5}N_{0}B^{(p+1)/2}
	\left(
	\frac{\nu}{2c_{1}}\right)^{-(p-1)/2} {\rm erg\,s}^{-1}{\rm Hz}^{-1},
\label{eq:Lthin1}
\end{equation}
and the synchrotron-self-absorbed luminosity is
\begin{equation}
	L_{\nu}=4\pi^{2}fR_{s}^{2}\frac{c_{5}}{c_{6}}B^{-1/2}
	\left(
	\frac{\nu}{2c_{1}}\right)^{5/2}  {\rm erg\,s}^{-1}{\rm Hz}^{-1},
\label{eq:Lssa}
\end{equation}
where $R_{s}=v_s t$ is the shock-wave radius, $B$ is the strength
of the magnetic field; and the constants $c_1$, $c_5$, and $c_6$
can be found in \cite{pacholczyk70}.  $N_0$ can be straightforwardly
shown to be $\epsilon_e/(8\pi\epsilon_B)B^2(p-2)(\gamma_{\rm min}m_ec^2)^{p-2}$.
Equation~\ref{eq:Lthin1} can be simplified to yield
\begin{equation}
	L_{\nu}=3\times 10^{34} \frac{128\pi^{3}f
	v_{s}^{3}}{3}c_{5}\gamma_{\rm min}m_{e}c^{2}\epsilon_{\rm e}\epsilon_{\rm B}
	\left(\frac{\dot M [M_{\odot} {\rm yr}^{-1}]}
	{w [{\rm km}~{\rm s}^{-1}]}\right)^{2}
	\left(\frac{\nu}{2c_{1}}\right)^{-1}t_{\rm d}^{-1}  {\rm erg\,s}^{-1}{\rm Hz}^{-1}.
\label{eq:Lthin3}
\end{equation}

Given that the earliest observation(s) were undertaken only a day after the
explosion it is prudent to check if there is significant free-free
absorption. The free-free optical depth is
\begin{equation}
	\tau_{\rm ff}=3.3\times 10^{-7}T_4^{-1.35}\nu_{\rm GHz}^{-2.1} {\rm EM}
\label{eq:tauff1}
\end{equation}
where $T=10^4T_4$ is the electron temperature (in degrees
Kelvin)\footnote{We use the convention of $X_n=X/10^n$ where it is
assumed, unless explicitly specified, that the units are CGS.},
$\nu_{\rm GHz}$ is the frequency in GHz and EM is the emission
measure, the integral of $n_e^2$ along the line of sight, and in
units of cm$^{-6}$\,pc.  The emission measure from a radius, say,
$r_*$ to infinity is
	\begin{eqnarray}
	{\rm EM} &=& \int_{r_*}^{\infty} n_*^2\Big( \frac{r}{r_*}\Big)^{-4}dr
	=\frac{1}{3}n_*^2r_*
	\end{eqnarray}
where $n_*$ is the density of electrons at radius $r_*$.  
Putting these equations together the free-free optical depth is 
\begin{equation}
	\tau_{\rm ff}\approx 0.5 \dot M_{-8}^2 t_{\rm d}^{-3} v_9^{-3} w_7^{-2} 
	T_4^{-1.35}\nu_{\rm GHz}^{-2.1}.
\end{equation}

For PTF11kly, the photospheric velocity is $v=2\times 10^9\,$cm\,s$^{-1}$.
The blast wave will at least have this velocity and likely
twice this value \citep{chevalier+fransson06, fryer+07, soderberg+10}. For representative values ($\dot M=3\times
10^{-7}\,M_\odot\,{\rm yr}^{-1}$, $w_7=1$, $v_9=4$) we find the
following optical depth at the first epoch of our observations:
$\tau_{\rm ff}[\nu_{\rm GHz}=1.4, 5, 8, 95]=[3.5, 0.24, 0.09, 0]$.
For interesting values of $\dot M$ ($\simlt 10^{-8}\,M_\odot\,{\rm yr}^{-1}$)
free-free optical depth is not important for the observations reported
here.

Equation~\ref{eq:Lthin3} provides the starting point for the discussion.
This equation shows that the spectral luminosity can constrain
$\dot M/w$ provided that we have a good grasp of the blast wave
dynamics and microphysics of particle acceleration and magnetic field
generation. We have argued above that $p\approx 3$, $\epsilon_e\approx 0.1$
and $v_s\approx 4\times 10^9\,$cm\,s$^{-1}$. We adopt these values and
proceed with the discussion.

\begin{figure}
\centering
\includegraphics[scale=0.4]{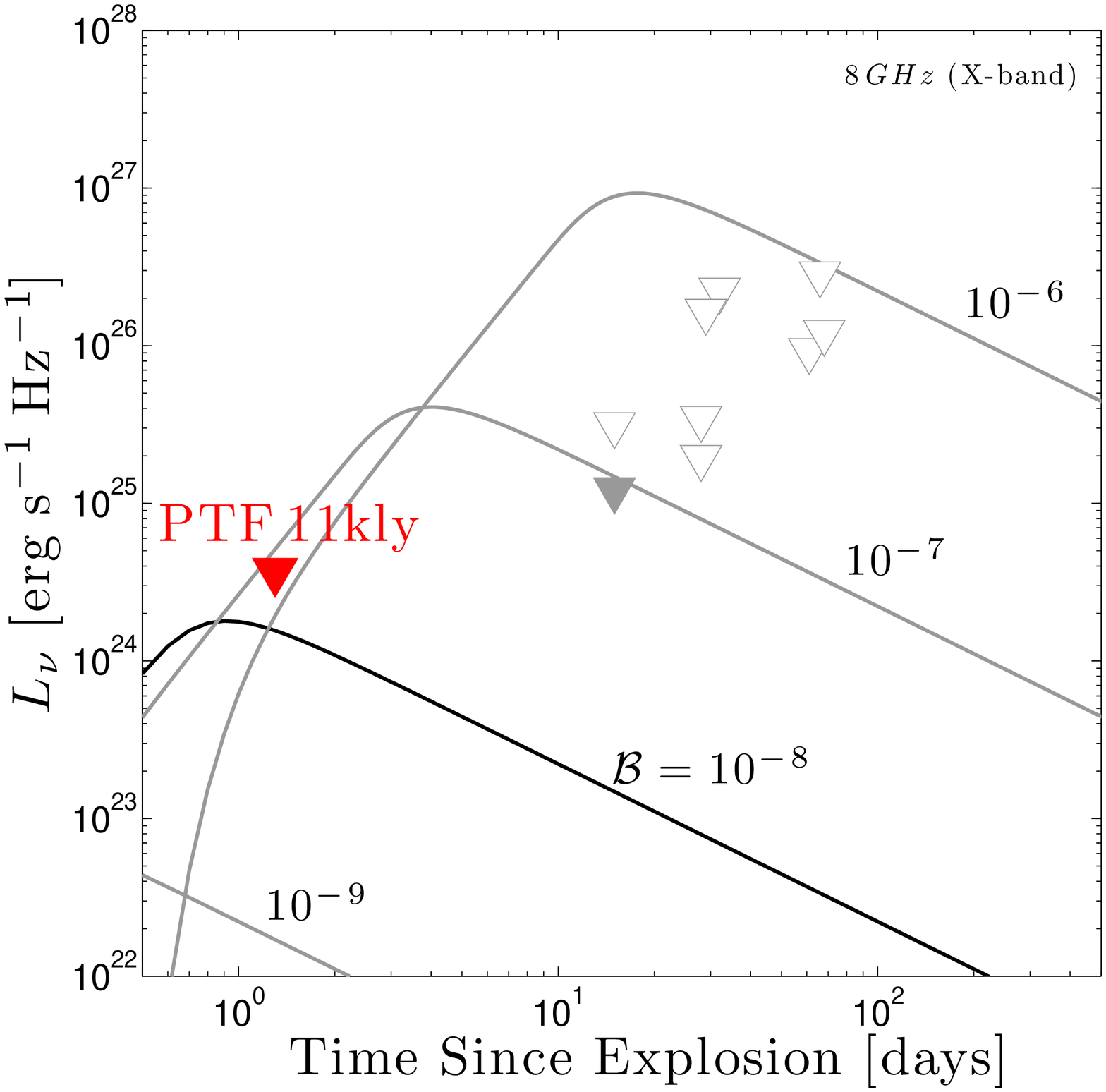}
\includegraphics[scale=0.4]{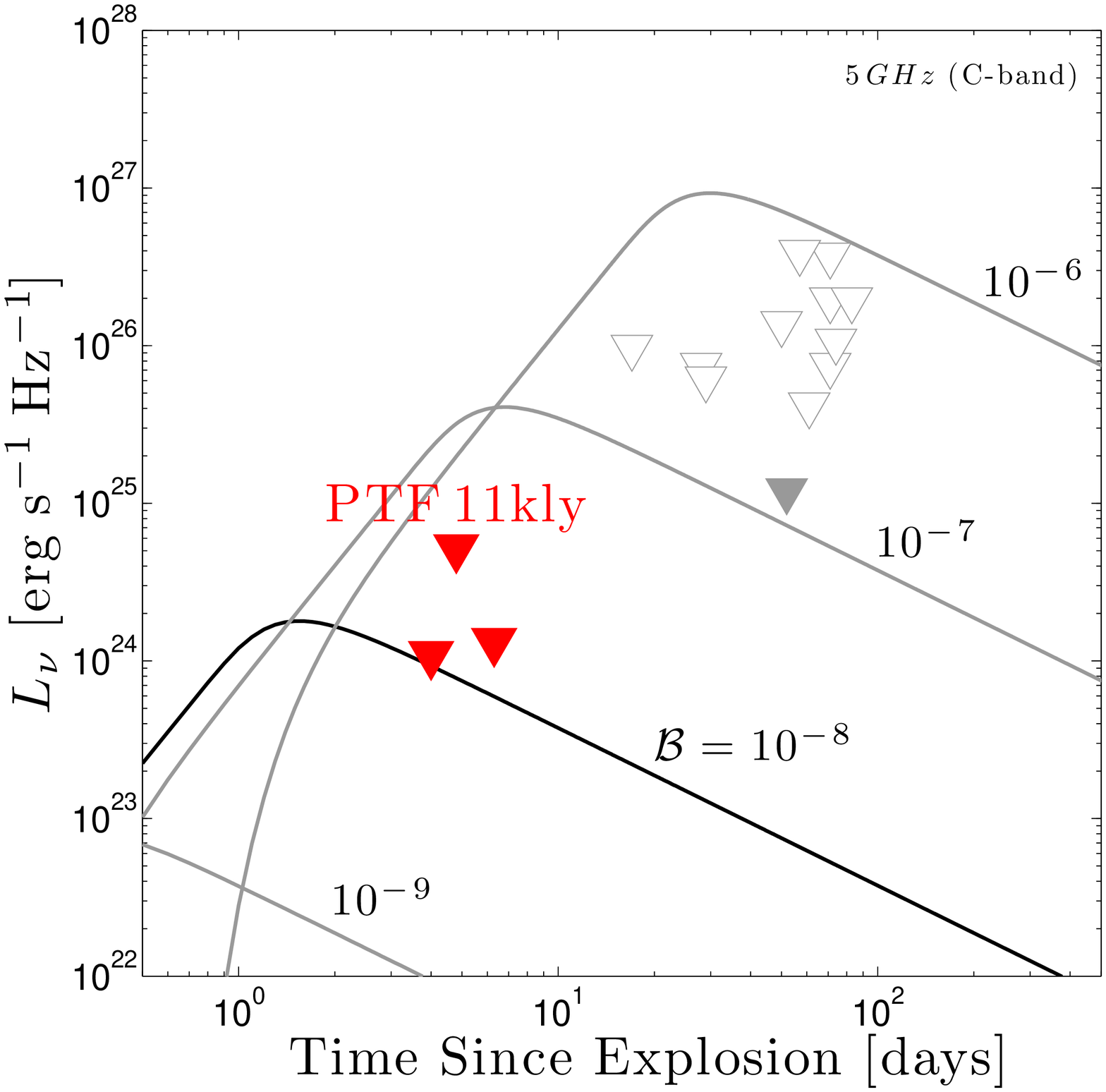}
\caption{\small
The model light curves for four different values of 
$\mathcal{B}=\epsilon \dot M/w$ where $\epsilon^2=\epsilon_B\epsilon_e$,
$w$ is the wind velocity and $\dot M$ is the mass loss rate (see also 
\S\ref{sec:ConstraintsfromRadio}).
We assume the following normalization, $\epsilon=0.1$,
$w=10^7\,{\rm cm\,s}^{-1}$ and $v_s=4\times 10^9\,{\rm cm\,s}^{-1}$. 
Model light curves for four values of $\mathcal{B}=[10^{-6}, 10^{-7},
10^{-8}, 10^{-9}] (0.1/\epsilon)w_{7}\,M_{\odot}{\rm yr}^{-1}$ are shown.
Top panel is for X
band (8\,GHz) while bottom panel is for C (5\,GHz) band. The open 
triangles in both panels show upper limits from
\citet{panagia+06}. The gray triangle in the top panel is the best
limit from \citet{panagia+06} while that in the bottom panel is the
upper limit from \citet{hancock+11}. Red triangles are measurements
of PTF11kly and presented in Table~\ref{tab:RadioLog}.
}
\label{fig:lc}
\end{figure}

To start with we can see that the observations reported here
(Table~\ref{tab:RadioLog}) yield
the lowest limits on the radio luminosity of very young Ia supernovae,
$L_\nu \simlt 10^{24}{\rm erg}~{\rm s}^{-1}{\rm Hz}^{-1}$.  
This limit then constrains the following parameter,
$\mathcal{B}\equiv\epsilon\dot M/w$, where
$\epsilon\equiv\sqrt{\epsilon_{\rm B}\epsilon_{\rm e}}$. 
We have deliberately not quoted the limits on $\dot M$ from previous
literature since radio measurements yield not $\dot M$ but $\mathcal{B}$
and this quantity depends on the unknown parameter, $\epsilon_B$ 
which (as summarized above) can vary by orders of magnitude. Despite this it is clear that
the observations reported here present the most sensitive limits to $\dot M$ to date
(Figure~\ref{fig:lc}).

From  Figure~\ref{fig:lc} we note that $\mathcal{B}$ can be constrained
as follows: the optically thin regime offers a lower bound whereas
the optically thick regime (SSA and free-free) an upper bound.   
The upper bound is not interesting since the simplest explanation for
the absence of radio emission is that the explosion takes place in a vacuum
(or very low density circumstellar medium). Besides, from past
studies, there is no indication of $\dot M$ in the range of $10^{-6}\,M_\odot\,{\rm yr}^{-1}$ that
is indicated from the upper bounds.

The constraint deduced for each
observation are summarized in Table~\ref{tab:RadioLog}. Combining
all the constraints we find $\dot M\simlt 1\times
10^{-8} w_7(0.1/\epsilon)M_{\odot} {\rm yr}^{-1}$.

\subsection{Constraints from X-ray data}
\label{sec:ConstraintsfromX-ray}

Past X-ray observations, typically undertaken no earlier than a
week past the explosion, have resulted in upper limits at the level
of $L_X\simlt 10^{39}\,{\rm erg\,s}^{-1}$.
A claim of detection of emission from SN\,2005ke ($L_X\sim 2\times
10^{38}{\rm erg\,s}^{-1}$; \citealt{immler+06}) has been disputed by
\citet{hughes+07}.  \citet{immler11} reported a detection in a
2-month stack of {\it Swift} XRT data of SN\,2011by. However, subsequent high
angular resolution Chandra observations by  \citet{pooley11} strongly
suggest that the emission arose from a steady source 2.3-arcsecond
away from the supernova.  The most sensitive and constraining
observation (prior to this event) was from a 20-ksec Chandra
observation of SN\,2002bo, obtained 9 days past the explosion
\citep{hughes+07}. The  3-$\sigma$ upper limit was $L_X <
1.3\times 10^{38}\,{\rm erg\,s}^{-1}$ (2--10 keV).

The {\it Swift} and Chandra X-ray observations of PTF\,11kly provide the earliest (1
day) as well as the most sensitive limits ($\approx$30 times deeper
relative to previous upper limits) on X-ray emission from SNe Ia.
As noted earlier (\S\ref{sec:MassLossRateConstraints})
there are two durable sources of X-ray emission: thermal
bremsstrahlung and inverse Compton (IC) scattering of the SN photons by
relativistic electrons.  For such strong shocks, kT=10\,keV is a
characteristic temperature (bremsstrahlung).  IC scattering from
1\,eV to the 0.1--10\,keV band requires electrons with Lorentz factor
$\gamma$ in the range of 10 to 100, a reasonable range for strong
(non-relativistic) shocks.

We first consider a model in which we ascribe all X-ray emission
(if any) to thermal bremsstrahlung.  Following the model discussed
in \citet{immler+06}, $L_X = (\pi m^2)^{-1}\, \Lambda(T)\,
(\dot{M}/w^2\, (v t)^{-1}$, and with the assumptions of an H+He
plasma as in \S\ref{sec:Observations}, $L_{\rm reverse}=30\, L_{\rm
forward}$, and $\Lambda(T) = 3\times 10^{-23}$\,erg cm$^3$ s$^{-1}$,
the upper limit of $\L_X< 4.4\times 10^{36}\,{\rm erg\,s}^{-1}$
translates into a progenitor mass-loss constraint of $\dot{M} \simlt
5\times 10^{-6}w_7\,M_\odot\,$yr$^{-1}$.  Scaling from the upper
limit for SN\,2002bo, given the distinct and more conservative model
assumptions of \citet{hughes+07},  suggests a limit for PTF\,11kly of
$\dot{M} \simlt 4\times 10^{-5}w_7\,M_\odot {\rm yr}^{-1}$.

Next, we consider the IC scattering model. From
\citet{chevalier+fransson06} we find that the IC luminosity is 
\begin{equation}
	L_{\rm IC} \approx 10^{36}\gamma_{\rm min}\Big(\frac{\epsilon_e}{0.1}\Big) v_9t_{\rm d}^{-1}
	\dot M_{-8}w_7^{-1} \Big(\frac{L_{\rm SN}}{10^{42}\,{\rm
	erg\,s}^{-1}}\Big) \,{\rm erg\,s}^{-1}.  
\label{eq:IC} 
\end{equation}
where $L_{\rm SN}$ is the bolometric luminosity from the supernova.
Clearly, the IC luminosity is independent of the highly variable
$\epsilon_B$ and, unlike the radio luminosity (Equation~\ref{eq:Lthin3})
has a gentle dependence on $v_{s}$. Assuming $\epsilon_e=0.1$ 
and noting that $L_{\rm SN}\sim 2\times10^{42}\,{\rm erg\,s}^{-1}$ on 
day 4 we find $\dot
M\simlt 10^{-8}w_7\,M_\odot\,{\rm yr}^{-1}$ (90\% confidence limit).  Unlike the radio
case this is a robust limit since it does not depend on $\epsilon_B$.

\section{Discussion}

As noted in \S\ref{sec:Introduction} the commonly accepted SN Ia
model is a thermonuclear explosion of a Carbon+Oxygen white dwarf
close to the Chandrasekhar mass. However, such white dwarfs
have masses smaller than the Chandrasekhar mass and so the Ia models
necessarily involve the growth of mass of the white dwarf.
One such model is the coalescence of two white dwarfs whose
total mass is in the vicinity of the Chandrasekhar mass (see
\citealt{vanKerkwijk10} and references therein).
There is no expectation of an enriched circumstellar medium 
for (long lived) double degenerate (DD) systems. Thus the
blast wave from the supernova is not expected to produce
any strong emission either in radio or X-ray bands.

On the other hand, in the ``Single Degenerate'' (SD) model the
white dwarf can grow by a steady transfer of matter from a 
companion. This can occur from a close binary in Roche Lobe Overflow
(RLOF), or by accretion directly from the wind of a companion (the
symbiotic channel). In this class of models one may expect
a circumstellar medium enriched by matter from the donor
star either directly (by a wind from the donor star) or indirectly
(matter that the white dwarf is unable to accrete is expelled
from the binary system).

It has long been the hope that early and sensitive radio and X-ray
observations of nearby Ia supernovae would diagnose the close in
circumstellar matter and thereby discriminate between the SD
and DD models.

Here, we report the earliest radio and X-ray observations of
PTF11kly and sensitive radio and X-ray observations undertaken
by us and others at premier radio and X-ray facilities. Despite
the sensitivity and rapidity no signal is seen at either radio
or X-ray bands. 

The absence of the signal is consistent with the expectation
of the DD model but would hardly constitute proof of this channel.
Below, we confront popular SD models with
our null detection.

The radio observations find that $\dot M\simlt 10^{-8}w_7(0.1/\epsilon)\,
M_{\odot}\,{\rm yr}^{-1}$, where $w= 10^7\,w_7\,{\rm cm\,s}^{-1}$
is the velocity with which the matter is ejected from the binary
system and $\epsilon=\sqrt{\epsilon_{\rm B}\epsilon_{e}}$. The X-ray observations yield an upper limit of 
$10^{-8}w_7^{-1}(0.1/\epsilon_{e})\,M_{\odot}\,{\rm yr}^{-1}$. To our knowledge, these
are the most sensitive limits on $\dot M$ reported for any
Ia supernova, to date.

Returning to the SD model, 
the growth in the mass of the white dwarf is not assured.  To start
with, the mass transfer rate has to be high enough to prevent a nova
explosion carrying off all the accreted matter. At high
enough rates, steady burning can occur on the white dwarf and the
white dwarf can grow in mass.  Accretion at a rate of $\approx
10^{-7}\,M_{\odot}\,{\rm yr}^{-1}$ is thought to be necessary for
stable accretion and nuclear burning on the surface of a white dwarf
(\citealt{Nomoto82}).  

Systems in RLOF may achieve the required accretion rate -- the class
of binaries known as 'supersoft x-ray sources' (van de Heuvel et
al. 1992 \nocite{VanDenHeuvel92}) show such behavior.  But in addition to steady accreation, mass loss from the system is also thought to occur.  Figure 1 from \citet{HanPodsiadlowski04} gives theoretically expected mass-loss rates for the classical supersoft channel using the WD accretion efficiencies from \citet{Hachisu99}.
In this phase, the mass loss from
the system can reach $10^{-6}\,M_\odot\,{\rm yr}^{-1}$ and, in extreme (but rare)
cases, more than $10^{-5}\,M_\odot\,{\rm yr}^{-1}$. 
The velocity of this material
depends on the details of the ejection process; if it comes from the
neighborhood of the WD, it is expected to be several
$10^8\,{\rm cm\,s}^{-1}$; if it comes from a circum-binary envelope, its
velocity should be at least of order the typical orbital velocity,
i.e.\ a few $10^7\,{\rm cm\,s}^{-1}$. Note, however, that in the vast
majority of cases, the binary will not be in this wind phase at the
time of explosion, and only a fraction of the mass transferred may be
lost from the system at that time; unfortunately, the exact amount is
not well constrained by the theoretical models (perhaps a few
$10^{-8}\,M_\odot$ with a velocity of at least a few
$10^{7}$\,cm\,s$^{-1}$). 

The radio limits for PTF\,11kly do not rule out
mass loss at that level. 
If the donor star is a main-sequence star or a slightly evolved sub-giant, its stellar wind is unlikely to be an important contribution to the systemic mass loss.

In the `symbiotic' SN Ia channel, the stellar wind from the donor star
may be more important than mass loss associated with the mass-transfer
process. If the companion is a Mira variable (in a D-type symbiotic),
the expected mass loss is $10^{-6}$ to $10^{-4}\,M_\odot\,{\rm yr}^{-1}$ 
(e.g.,
\citealt{Zijlstra06}) with very low terminal velocity ($w\sim 10\,{\rm km\,s^{-1}}$ or
even less). Such high mass loss can clearly be ruled out in the case
of PTF 11kly. 

In an S-type symbiotic, which is a more likely SN Ia
progenitor (\citealt{Hachisu99}), the donor star is a much less
evolved red giant with much lower mass loss.  \citet{Seaquist90} estimate typical rates of 
$10^{-8}$--$10^{-7}\, M_\odot\,{\rm yr}^{-1}$ with
large modeling uncertainties. A case of particular interest is the
symbiotic binary EG And which contains a M2.4 red giant that is very
similar to the red giant in the SN Ia candidate system RS Oph. The wind
from EG And has a measured high terminal velocity of 75\,km\,s$^{1}$
(\citealt{EspeyCrowley06}) with an uncertain estimate of the mass loss rate of
about $10^{-8}\,M_\odot\,{\rm yr}^{-1}$ (\citealt{Crowley06}). For comparison, the observed
wind features in RS Oph have velocities in the range of
30--60\,km\,s$^{-1}$ (\citealt{Patat11}). In addition, observations
of RS Oph show that the mass loss is very asymmetric and is strongly
confined to the orbital plane (\citealt{O'Brien06}), consistent with
hydrodynamical modeling of the mass loss from such systems (\citealt{Walder08}; Mohamed, Booth \& Podsiadlowski [in
  preparation]). This introduces a possibly important viewing
dependence of the radio signal.  In this context, it is worth noting
that these systems (both in the supersoft and the symbiotic channel)
experience recurrent novae, typically every 10--20\,yr when the white
dwarf is close to the Chandrasekhar mass (\citealt{Schaefer10}), which may
produce low-density cavities in the immediate circumstellar medium
(\citealt{Wood-Vasey06}).  The upper limit for the mass-loss
rate deduced from our radio limits for the progenitor of PTF 11kly is
comparable to the expected mass loss from a symbiotic like RS Oph.
However, considering the uncertainties noted above, it may be
premature to rule out such a system on the basis of these observations
alone.

The limits we have derived from radio and X-ray observations on the
progenitor system of PTF11kly can be placed into context with complementary
constraints from a variety of independent techniques.  \citet{lbp+11}
examine pre-explosion \textit{HST} images of M101 and derive limits on the
brightness of any mass-donating companion.  The lack of any optical 
emission at the location of PTF11kly directly rules out luminous red giants
and the vast majority of He stars, consistent with the limits we have
derived here. 

In summary, we present the most sensitive early radio and X-ray
observations to date. Despite the rapid response and excellent
sensitivity our observations can constrain only the symbiotic SD channel
for PTF11kly, but cannot rule out a main-sequence or sub-giant donor channel. 

Prior observations which were both less sensitive and at later
times nominally came to the same conclusion. However, the inference
of $\dot M$ from radio data depends on two microphysics parameters,
$\epsilon_e$, $\epsilon_B$ and two velocities, $v_s$ (shock
speed) and 
$w$, the 
velocity with which matter is ejected from the binary system.
In principle, $v_s$ can be estimated from photospheric velocities
and a model for the exploding white dwarf. The value of $\epsilon_e$
empirically shows modest variation and one can assume $\epsilon_e\approx
0.1$. However, $\epsilon_B$, on empirical and theoretical grounds can
vary tremendously (with $\epsilon_B\approx 0.1$ being a maximum
plausible value).

In the past, particularly in the radio literature,
the assumed values were (in our opinion) rather
optimistic: $w\sim 10^6\,{\rm cm\,s}^{-1}$ and $\epsilon_B=0.1$.
Using the maximum possible value for $\epsilon_B\approx 0.1$
we find from our radio observations that $\dot M\simlt 10^{-8}w_7
\,M_{\odot}\,{\rm yr}^{-1}$. Smaller values of $\epsilon_B$ will only
make this limit worse (proportionally larger).
Given that PTF11kly was one of the closest Ia supernovae it is 
not likely that the limits presented here would be bested
in the near term.

The X-ray observations provide a less model dependent
(and hence more robust)
estimate of $\dot M$. The X-ray observations (especially if undertaken
at peak) tightly constrain $\dot M\epsilon_e/w$.  The microphysical
parameter, $\epsilon_e$ has far less dispersion as compared to 
$\epsilon_B$ and as such the X-ray observations yield a robust
estimate of $\dot M$ (relative to that obtained from radio
observations). We find $\dot M\simlt 10^{-8}w_7\,M_{\odot}\,{\rm yr}^{-1}$.
The X-ray observations are quite sensitive and it is not likely that
these limits will be easily surpassed in this decade.

\section*{Acknowledgments}

We thank the CARMA and EVLA staff for promptly scheduling this
target of opportunity.  This work made use of data supplied by the
UK {\it Swift} Science Data Centre at the University of Leicester.
We thank the ASTRON Radio Observatory for the generous and swift allocation of observing time.
PTF is a fully-automated, wide-field survey aimed at a systematic exploration 
of explosions and variable phenomena in optical wavelengths. The participating
institutions are Caltech, Columbia University, Weizmann Institute of Science,
Lawrence Berkeley Laboratory, Oxford and University of California at Berkeley.
The program is 
centered on a 12Kx8K, 7.8 square degree CCD array (CFH12K) re-engineered for 
the 1.2-m Oschin Telescope at the Palomar Observatory by Caltech Optical Observatories.
Photometric follow-up is undertaken by the automated Palomar 1.5-m telescope. 
Research at Caltech is 
supported by grants from NSF and NASA. The Weizmann PTF partnership is supported in part by the Israeli Science Foundation
via grants to A.G. Weizmann-Caltech collaboration is supported by a grant from the BSF
to A.G. and S.R.K. A.G. further acknowledges the Lord Sieff of Brimpton Foundation. MMK acknowledges support from the Hubble fellowship
and Carnegie-Princeton Fellowship. We thank the ASTRON Radio Observatory for 
the generous and swift allocation of observing time. The Westerbork Synthesis Radio 
Telescope is operated by  ASTRON (Netherlands Foundation for Radio Astronomy) with 
support from the Netherlands Organization for Scientific Research (NWO). AJvdH was 
supported by NASA grant NNH07ZDA001-GLAST. 
S.B.C. acknowledges generous financial assistance from Gary \& Cynthia
Bengier, the Richard \& Rhoda Goldman Fund, NASA/{\it Swift} grants
NNX10AI21G and GO-7100028, the TABASGO Foundation, and NSF grant
AST-0908886. AK is partially supported by NSF award AST-1008353.

\bibliography{PTF11kly_ApJ_refereed2}

\begin{thebibliography}{54}
\expandafter\ifx\csname natexlab\endcsname\relax\def\natexlab#1{#1}\fi

\bibitem[{{Baars} {et~al.}(1977){Baars}, {Genzel}, {Pauliny-Toth}, \&
  {Witzel}}]{Baars77}
{Baars}, J.~W.~M., {Genzel}, R., {Pauliny-Toth}, I.~I.~K., \& {Witzel}, A.
  1977, \aap, 61, 99

\bibitem[{{Bj{\"o}rnsson} \& {Fransson}(2004)}]{Bjornsson2004}
{Bj{\"o}rnsson}, C.-I., \& {Fransson}, C. 2004, \apj, 605, 823

\bibitem[{{Boffi} \& {Branch}(1995)}]{Boffi+branch95}
{Boffi}, F.~R., \& {Branch}, D. 1995, \pasp, 107, 347

\bibitem[{{Branch} {et~al.}(1995){Branch}, {Livio}, {Yungelson}, {Boffi}, \&
  {Baron}}]{boffi+95}
{Branch}, D., {Livio}, M., {Yungelson}, L.~R., {Boffi}, F.~R., \& {Baron}, E.
  1995, \pasp, 107, 1019

\bibitem[{{Chevalier}(1982)}]{chevalier82}
{Chevalier}, R.~A. 1982, \apj, 259, 302

\bibitem[{{Chevalier}(1998)}]{chevalier98}
---. 1998, \apj, 499, 810

\bibitem[{{Chevalier} \& {Fransson}(2006)}]{chevalier+fransson06}
{Chevalier}, R.~A., \& {Fransson}, C. 2006, \apj, 651, 381

\bibitem[{{Chomiuk} \& {Soderberg}(2011)}]{chomiuk+soderberg11}
{Chomiuk}, L., \& {Soderberg}, A. 2011, The Astronomer's Telegram, 3532, 1

\bibitem[{{Crowley}(2006)}]{Crowley06}
{Crowley}, C. 2006, PhD thesis, School of Physics, Trinity College Dublin,
  Dublin 2, Ireland Ireland

\bibitem[{{Eck} {et~al.}(1995){Eck}, {Cowan}, {Roberts}, {Boffi}, \&
  {Branch}}]{eck+95}
{Eck}, C.~R., {Cowan}, J.~J., {Roberts}, D.~A., {Boffi}, F.~R., \& {Branch}, D.
  1995, \apjl, 451, L53+

\bibitem[{{Espey} \& {Crowley}(2008)}]{EspeyCrowley06}
{Espey}, B.~R., \& {Crowley}, C. 2008, in Astronomical Society of the Pacific
  Conference Series, Vol. 401, RS Ophiuchi (2006) and the Recurrent Nova
  Phenomenon, ed. {A.~Evans, M.~F.~Bode, T.~J.~O'Brien, \& M.~J.~Darnley},
  166--+

\bibitem[{{Evans} {et~al.}(2009){Evans}, {Beardmore}, {Page}, {Osborne},
  {O'Brien}, {Willingale}, {Starling}, {Burrows}, {Godet}, {Vetere}, {Racusin},
  {Goad}, {Wiersema}, {Angelini}, {Capalbi}, {Chincarini}, {Gehrels}, {Kennea},
  {Margutti}, {Morris}, {Mountford}, {Pagani}, {Perri}, {Romano}, \&
  {Tanvir}}]{evans+09}
{Evans}, P.~A., {et~al.} 2009, \mnras, 397, 1177

\bibitem[{{Feigelson} {et~al.}(2002){Feigelson}, {Broos}, {Gaffney}, {Garmire},
  {Hillenbrand}, {Pravdo}, {Townsley}, \& {Tsuboi}}]{feigelson+02}
{Feigelson}, E.~D., {Broos}, P., {Gaffney}, III, J.~A., {Garmire}, G.,
  {Hillenbrand}, L.~A., {Pravdo}, S.~H., {Townsley}, L., \& {Tsuboi}, Y. 2002,
  \apj, 574, 258

\bibitem[{{Fryer} {et~al.}(2007){Fryer}, {Mazzali}, {Prochaska}, {Cappellaro},
  {Panaitescu}, {Berger}, {van Putten}, {van den Heuvel}, {Young},
  {Hungerford}, {Rockefeller}, {Yoon}, {Podsiadlowski}, {Nomoto}, {Chevalier},
  {Schmidt}, \& {Kulkarni}}]{fryer+07}
{Fryer}, C.~L., {et~al.} 2007, \pasp, 119, 1211

\bibitem[{{Gal-Yam} {et~al.}(2011){Gal-Yam}, {Kasliwal}, {Arcavi}, {Green},
  {Yaron}, {Ben-Ami}, {Xu}, {Sternberg}, {Quimby}, {Kulkarni}, {Ofek},
  {Walters}, {Nugent}, {Poznanski}, {Bloom}, {Cenko}, {Filippenko}, {Li},
  {Silverman}, {Walker}, {Sullivan}, {Maguire}, {Howell}, {Mazzali}, {Frail},
  {Bersier}, {James}, {Akerlof}, {Yuan}, {Law}, {Fox}, \&
  {Gehrels}}]{Gal-Yam11}
{Gal-Yam}, A., {et~al.} 2011, \apj, 736, 159

\bibitem[{{Hachisu} {et~al.}(1999){Hachisu}, {Kato}, \& {Nomoto}}]{Hachisu99}
{Hachisu}, I., {Kato}, M., \& {Nomoto}, K. 1999, \apj, 522, 487

\bibitem[{{Han} \& {Podsiadlowski}(2004)}]{HanPodsiadlowski04}
{Han}, Z., \& {Podsiadlowski}, P. 2004, \mnras, 350, 1301

\bibitem[{{Hancock} {et~al.}(2011){Hancock}, {Gaensler}, \&
  {Murphy}}]{hancock+11}
{Hancock}, P.~P., {Gaensler}, B.~M., \& {Murphy}, T. 2011, \apjl, 735, L35+

\bibitem[{{Hillebrandt} \& {Niemeyer}(2000)}]{hillebrandt+niemeyer00}
{Hillebrandt}, W., \& {Niemeyer}, J.~C. 2000, \araa, 38, 191

\bibitem[{Horesh(2011)}]{Horesh2011}
Horesh, A. 2011, in prep.

\bibitem[{{Hoyle} \& {Fowler}(1960)}]{hoyle+fowler60}
{Hoyle}, F., \& {Fowler}, W.~A. 1960, \apj, 132, 565

\bibitem[{{Hughes} {et~al.}(2007){Hughes}, {Chugai}, {Chevalier}, {Lundqvist},
  \& {Schlegel}}]{hughes+07}
{Hughes}, J.~P., {Chugai}, N., {Chevalier}, R., {Lundqvist}, P., \& {Schlegel},
  E. 2007, \apj, 670, 1260

\bibitem[{{Hughes} {et~al.}(2011){Hughes}, {Soderberg}, \& {Slane}}]{Hughes+11}
{Hughes}, J.~P., {Soderberg}, A., \& {Slane}, P. 2011, The Astronomer's
  Telegram, 3602, 1

\bibitem[{{Iben} \& {Tutukov}(1984)}]{iben+tutkov84}
{Iben}, Jr., I., \& {Tutukov}, A.~V. 1984, \apjs, 54, 335

\bibitem[{{Immler} \& {Russell}(2011)}]{immler11}
{Immler}, S., \& {Russell}, B.~R. 2011, The Astronomer's Telegram, 3410, 1

\bibitem[{{Immler} {et~al.}(2006){Immler}, {Brown}, {Milne}, {The}, {Petre},
  {Gehrels}, {Burrows}, {Nousek}, {Williams}, {Pian}, {Mazzali}, {Nomoto},
  {Chevalier}, {Mangano}, {Holland}, {Roming}, {Greiner}, \&
  {Pooley}}]{immler+06}
{Immler}, S., {et~al.} 2006, \apjl, 648, L119

\bibitem[{{Kalberla} {et~al.}(2005){Kalberla}, {Burton}, {Hartmann}, {Arnal},
  {Bajaja}, {Morras}, \& {P{\"o}ppel}}]{kalberla+05}
{Kalberla}, P.~M.~W., {Burton}, W.~B., {Hartmann}, D., {Arnal}, E.~M.,
  {Bajaja}, E., {Morras}, R., \& {P{\"o}ppel}, W.~G.~L. 2005, \aap, 440, 775

\bibitem[{{Kamphuis} {et~al.}(1991){Kamphuis}, {Sancisi}, \& {van der
  Hulst}}]{Kamphuis1991}
{Kamphuis}, J., {Sancisi}, R., \& {van der Hulst}, T. 1991, \aap, 244, L29

\bibitem[{{Law} {et~al.}(2009){Law}, {Kulkarni}, {Dekany}, {Ofek}, {Quimby},
  {Nugent}, {Surace}, {Grillmair}, {Bloom}, {Kasliwal}, {Bildsten}, {Brown},
  {Cenko}, {Ciardi}, {Croner}, {Djorgovski}, {van Eyken}, {Filippenko}, {Fox},
  {Gal-Yam}, {Hale}, {Hamam}, {Helou}, {Henning}, {Howell}, {Jacobsen},
  {Laher}, {Mattingly}, {McKenna}, {Pickles}, {Poznanski}, {Rahmer}, {Rau},
  {Rosing}, {Shara}, {Smith}, {Starr}, {Sullivan}, {Velur}, {Walters}, \&
  {Zolkower}}]{law+09}
{Law}, N.~M., {et~al.} 2009, \pasp, 121, 1395

\bibitem[{{Li} {et~al.}(2011){Li}, {Bloom}, {Podsiadlowski}, {Miller}, {Cenko},
  {Jha}, {Sullivan}, {Howell}, {Nugent}, {Butler}, {Ofek}, {Kasliwal},
  {Richards}, {Stockton}, {Shih}, {Bildsten}, {Shara}, {Bibby}, {Filippenko},
  {Ganeshalingam}, {Silverman}, {Kulkarni}, {Law}, {Poznanski}, {Quimby},
  {McCully}, {Patel}, \& {Maguire}}]{lbp+11}
{Li}, W., {et~al.} 2011, submitted to \textit{Nature}\ (astro-ph/1109.1593)

\bibitem[{{Nomoto}(1982)}]{Nomoto82}
{Nomoto}, K. 1982, \apj, 253, 798

\bibitem[{{Nugent} {et~al.}(2011)}]{Nugent11}
{Nugent}, P.~E., {et~al.} 2011, submitted to \textit{Nature}

\bibitem[{{O'Brien} {et~al.}(2006){O'Brien}, {Bode}, {Porcas}, {Muxlow},
  {Eyres}, {Beswick}, {Garrington}, {Davis}, \& {Evans}}]{O'Brien06}
{O'Brien}, T.~J., {et~al.} 2006, \nat, 442, 279

\bibitem[{{Pacholczyk}(1970)}]{pacholczyk70}
{Pacholczyk}, A.~G. 1970, {Radio astrophysics. Nonthermal processes in galactic
  and extragalactic sources}, ed. {Pacholczyk, A G.}

\bibitem[{{Panagia} {et~al.}(2006){Panagia}, {Van Dyk}, {Weiler}, {Sramek},
  {Stockdale}, \& {Murata}}]{panagia+06}
{Panagia}, N., {Van Dyk}, S.~D., {Weiler}, K.~W., {Sramek}, R.~A., {Stockdale},
  C.~J., \& {Murata}, K.~P. 2006, \apj, 646, 369

\bibitem[{{Patat} {et~al.}(2011){Patat}, {Chugai}, {Podsiadlowski}, {Mason},
  {Melo}, \& {Pasquini}}]{Patat11}
{Patat}, F., {Chugai}, N.~N., {Podsiadlowski}, P., {Mason}, E., {Melo}, C., \&
  {Pasquini}, L. 2011, \aap, 530, A63+

\bibitem[{{Perlmutter} {et~al.}(1999){Perlmutter}, {Aldering}, {Goldhaber},
  {Knop}, {Nugent}, {Castro}, {Deustua}, {Fabbro}, {Goobar}, {Groom}, {Hook},
  {Kim}, {Kim}, {Lee}, {Nunes}, {Pain}, {Pennypacker}, {Quimby}, {Lidman},
  {Ellis}, {Irwin}, {McMahon}, {Ruiz-Lapuente}, {Walton}, {Schaefer}, {Boyle},
  {Filippenko}, {Matheson}, {Fruchter}, {Panagia}, {Newberg}, {Couch}, \& {The
  Supernova Cosmology Project}}]{perlmutter+99}
{Perlmutter}, S., {et~al.} 1999, \apj, 517, 565

\bibitem[{{Pooley}(2011)}]{pooley11}
{Pooley}, D. 2011, The Astronomer's Telegram, 3456, 1

\bibitem[{{Rau} {et~al.}(2009){Rau}, {Kulkarni}, {Law}, {Bloom}, {Ciardi},
  {Djorgovski}, {Fox}, {Gal-Yam}, {Grillmair}, {Kasliwal}, {Nugent}, {Ofek},
  {Quimby}, {Reach}, {Shara}, {Bildsten}, {Cenko}, {Drake}, {Filippenko},
  {Helfand}, {Helou}, {Howell}, {Poznanski}, \& {Sullivan}}]{rau+09}
{Rau}, A., {et~al.} 2009, \pasp, 121, 1334

\bibitem[{{Riess} {et~al.}(1998){Riess}, {Filippenko}, {Challis},
  {Clocchiatti}, {Diercks}, {Garnavich}, {Gilliland}, {Hogan}, {Jha},
  {Kirshner}, {Leibundgut}, {Phillips}, {Reiss}, {Schmidt}, {Schommer},
  {Smith}, {Spyromilio}, {Stubbs}, {Suntzeff}, \& {Tonry}}]{riess+98}
{Riess}, A.~G., {et~al.} 1998, \aj, 116, 1009

\bibitem[{{Schaefer}(2010)}]{Schaefer10}
{Schaefer}, B.~E. 2010, \apjs, 187, 275

\bibitem[{{Seaquist} \& {Taylor}(1990)}]{Seaquist90}
{Seaquist}, E.~R., \& {Taylor}, A.~R. 1990, \apj, 349, 313

\bibitem[{{Shappee} \& {Stanek}(2011)}]{shappee+stanek11}
{Shappee}, B.~J., \& {Stanek}, K.~Z. 2011, \apj, 733, 124

\bibitem[{{Soderberg} {et~al.}(2010){Soderberg}, {Brunthaler}, {Nakar},
  {Chevalier}, \& {Bietenholz}}]{soderberg+10}
{Soderberg}, A.~M., {Brunthaler}, A., {Nakar}, E., {Chevalier}, R.~A., \&
  {Bietenholz}, M.~F. 2010, \apj, 725, 922

\bibitem[{{Soderberg} {et~al.}(2011){Soderberg}, {Margutti}, {Zauderer},
  {Krauss}, {Katz}, {Chomiuk}, {Dittmann}, {Nakar}, {Sakamoto}, {Kawai},
  {Hurley}, {Barthelmy}, {Toizumi}, {Morii}, {Chevalier}, {Gurwell},
  {Petitpas}, {Rupen}, {Alexander}, {Levesque}, {Fransson}, {Brunthaler},
  {Bietenholz}, {Chugai}, {Connaughton}, {Briggs}, {Meegan}, {von Kienlin},
  {Zhang}, {Rau}, {Golenetskii}, {Mazets}, \& {Cline}}]{Soderberg+11}
{Soderberg}, A.~M., {et~al.} 2011, ArXiv e-prints

\bibitem[{{van den Heuvel} {et~al.}(1992){van den Heuvel}, {Bhattacharya},
  {Nomoto}, \& {Rappaport}}]{VanDenHeuvel92}
{van den Heuvel}, E.~P.~J., {Bhattacharya}, D., {Nomoto}, K., \& {Rappaport},
  S.~A. 1992, \aap, 262, 97

\bibitem[{{van Kerkwijk} {et~al.}(2010){van Kerkwijk}, {Chang}, \&
  {Justham}}]{vanKerkwijk10}
{van Kerkwijk}, M.~H., {Chang}, P., \& {Justham}, S. 2010, \apjl, 722, L157

\bibitem[{{Walder} {et~al.}(2008){Walder}, {Folini}, \& {Shore}}]{Walder08}
{Walder}, R., {Folini}, D., \& {Shore}, S.~N. 2008, \aap, 484, L9

\bibitem[{{Webbink}(1984)}]{webbink84}
{Webbink}, R.~F. 1984, \apj, 277, 355

\bibitem[{{Weiler} {et~al.}(2002){Weiler}, {Panagia}, {Montes}, \&
  {Sramek}}]{weiler+02}
{Weiler}, K.~W., {Panagia}, N., {Montes}, M.~J., \& {Sramek}, R.~A. 2002,
  \araa, 40, 387

\bibitem[{{Whelan} \& {Iben}(1973)}]{whelan+iben73}
{Whelan}, J., \& {Iben}, Jr., I. 1973, \apj, 186, 1007

\bibitem[{{Wood-Vasey} \& {Sokoloski}(2006)}]{Wood-Vasey06}
{Wood-Vasey}, W.~M., \& {Sokoloski}, J.~L. 2006, \apjl, 645, L53

\bibitem[{{Yungelson} \& {Livio}(2000)}]{yungelson+livio00}
{Yungelson}, L.~R., \& {Livio}, M. 2000, \apj, 528, 108

\bibitem[{{Zijlstra}(2006)}]{Zijlstra06}
{Zijlstra}, A.~A. 2006, in IAU Symposium, Vol. 234, Planetary Nebulae in our
  Galaxy and Beyond, ed. {M.~J.~Barlow \& R.~H.~M{\'e}ndez}, 55--62

\end{thebibliography}

\end{document}